\documentclass{article}

\usepackage{arxiv}

\usepackage[utf8]{inputenc} 
\usepackage[T1]{fontenc}    
\usepackage[hidelinks]{hyperref}      
\usepackage{url}            
\usepackage{booktabs}       
\usepackage{amsfonts}       
\usepackage{nicefrac}       
\usepackage{microtype}      
\usepackage{lipsum}		
\usepackage{graphicx}
\usepackage{natbib}
\usepackage{doi}
\usepackage{rotating}
\usepackage{subfig}
\usepackage{makecell}
\usepackage{xcolor}
\usepackage[normalem]{ulem}
\usepackage{soul}

\usepackage{floatrow}
\usepackage{caption}
\usepackage{subcaption}
\usepackage[rightcaption]{sidecap}

\title{CoARA will not save science from the tyranny of administrative evaluation}

\date{} 					

\author{\href{https://orcid.org/0000-0003-0293-482X}{\includegraphics[scale=0.06]{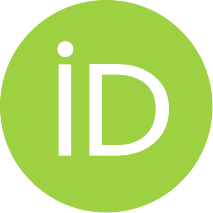}\hspace{1mm}Alberto Baccini}
  \\
	Dipartimento di Economia Politica e Statistica\\
	Università degli Studi di Siena\\
	Siena, Italy \\
	\texttt{alberto.baccini@unisi.it} \\
}

\renewcommand{\headeright}{}
\renewcommand{\undertitle}{}
\renewcommand{\shorttitle}{CoARA and the tyranny of administrative evaluation of research}

\hypersetup{
pdftitle={CoARA will not save us},
}

\makeatletter
\input{size12.clo}
\makeatother

\begin{document}

\maketitle

\begin{abstract}

The Coalition for Advancing Research Assessment (CoARA) agreement is a cornerstone in the ongoing efforts to reform research evaluation. CoARA advocates for administrative evaluations of research that rely on peer review, supported by responsible metrics, as beneficial for both science and society. Its principles can be critically examined through the lens of Philip Kitcher’s concept of well-ordered science in a democratic society. From Kitcher’s perspective, CoARA’s approach faces two significant challenges: definitions of quality and impact are determined by governments or evaluation institutions rather than emerging from broad public deliberation, and a select group of scientists is empowered to assess research based on these predefined criteria. This creates susceptibility to both the "tyranny of expertise" and the "tyranny of ignorance" that Kitcher cautions against. Achieving Kitcher’s ideal would require limiting administrative evaluations to essential tasks, such as researcher recruitment and project funding, while establishing procedures grounded in principles of fairness.

\end{abstract}


\newpage

\section{Introduction}

The agreement of the Coalition for Advancing Research Assessment (CoARA) is one of the key documents of the contemporary discussion on the role of research evaluation in the organization of science \citep{CoARA2024}. 
It comes after the so-called Leiden Manifesto on Research Assessment \citep{leiden}, after the advocacy for the responsible use of metrics in evaluation \citep{wilsdon}, and more than ten years after DORA, the San Francisco Declaration on Research Assessment \citep{dora}. 
All these documents, with others (for a syntesis see \citep{Rushfort}), can be considered as attempts by the ``science reform movement'' to respond to the long-lasting crisis of contemporary science \citep{Peterson2020MetascienceAA}.

All these documents are mainly focused on research evaluation technologies, i.e. on the techniques used for realizing evaluation. CoARA promotes the widespread use of peer-review technology supported by the responsible use of metrics, as defined in the Leiden manifesto \citep{leiden, wilsdon}. The use of peer review is instrumental in shifting the focus from traditional metrics, such as publication counts and citation indices, to a broader appreciation of diverse academic contributions, including teaching, leadership, openness, and societal impact. The debate between reformers and advocates of metrics is ongoing. In the Netherlands, for example, proponents of the Recognition and Rewards Program, considered the main ongoing experiment in assessment reform, face opposition from professors who express concern about the potential risks, especially for young scholars, arising from 'the abandonment of international evaluation criteria' \citep{Rushfort}. The tone of the debate is much more heated with regard to CoARA directly. The primary criticisms come from advocates of scientometrics \citep{abramo2024}, who even accuse CoARA of 'scientometric denialism' \citep{torres}. Critics did not spare even the narrative curriculum, one of the main instruments proposed by CoARA. It consists of a written narrative that contextualizes a researcher's contributions, explaining the quality, relevance, and impact of their work in a broader sense—including teaching, leadership, open science, and societal impact. \citet{TORRESSALINAS2024101546} ridiculized the narrative curriculum by launching the idea of ``narrative bibliometrics'', and illustrating how to write a narrative bibliometric curriculum, by using word to contextualize metrics . 

Reform documents and the related discussions relegate crucial issues to an invisible background such as the social desirability of evaluation, the costs and benefits associated with it, the institutional context within which it develops, and the fairness of the procedures that characterize it. In particular, they did not consider foundational questions such as the relations between research evaluation, democracy, academic freedom, pluralism, and the soundness of science (see, among others, \citep{gillies, Collini_2012,DeNicolao, Caso, pievatolo_2024} 

The question this paper asks is whether the generalized adoption of the CoARA principles could surely make the evaluation of research desirable and solve all its problems \citep{gillies, Caso, bpdn}. The answer to this question is no. To reach this conclusion, three main arguments are discussed. The first consists to define the notion of “administrative evaluation of research” as distinct from the evaluation activities inherent in science. The second characterizes CoARA as a form of `technocracy' perfectly consistent with a utilitarian view of research. The third consists in the adoption of an alternative view: Philip Kitcher's idea of well-ordered science. It is argued that administrative evaluation of research, even if correct according to CoARA principles, is at odds with the principles of well-ordered science. These three arguments allow to suggest some policy recipes, that in turn are at odds with current views about research evaluation.    

\section{Administrative evaluation of research}

CoARA, the Leiden Manifesto, DORA are three key documents of the evaluation reform movement. They are all concerned about the way in which research evaluation can be improved. But what kind of evaluation do these documents talk about? They are talking about \textit{administrative evaluation of research}. 

``Administrative evaluation'' generically indicates evaluation ``integrated as administrative routine at many levels and in many types of institutions'' \citep[p. 2]{dahler-larsen}. ``Administrative evaluation of research'' refers to evaluation carried out by governments, academic institutions, agencies,  and other organizations to produce indicators or other evaluation information. As a consequence, ``the so-called research evaluation has fully entered into administrative activities (selection, funding, etc.) and will eventually follow its rules. [...] Evaluation [adds] a stage to administrative proceedings of various kinds in which the ultimate decision-maker is the judge'' \citep{Cassese}.

Indicators and information obtained from administrative research evaluation are typically regarded as the fundamental components of accountability for the use of taxpayer funds in research. They also serve for evidence-based science policy and university governance \citep{Whitley}: they are instrumental in decisions regarding recruitment, promotions, and awards for individual researchers, as well as for project funding and performance-based research funding (PBRF) systems \citep{pbrf}. \citet{Molas} proposed to distinguish three main uses of the administrative research evaluation: distributive, aimed at distributing resources according to the results of evaluation;  improvement, aimed at individuating the best practice; and controlling, aimed at scritinizing the use of public resurces. 
As a matter of fact ``administrative evaluation of research'' encompasses massive research evaluation exercises such as the British Research Excellence Framework (REF), the Italian evaluation of research quality (VQR) or the Australian Excellence in Research (ERA); the performance based funding systems in the European nordic countries \citep{aagard}, massive evaluations conducted at the individual level as in the case of the Spanish ``sexenio'' \citep{marini} or the qualification for professorship (ASN) in Italy; evaluation of projects for funding as in the European Research Council competition for ERC grant. Indeed, research governance structures vary across countries, leading to diverse applications and uses of administrative research evaluation.

The espression ``administrative evaluation of research'' instead of ``research evaluation'' allows to carefully distinguish it from the evaluation that scientists continuously carried out during their work. When a scientist reads a paper or a book and decides to recommend it for publication, or to use it in their own research, they implicitly evaluate it as worthy of use because of its soundness, originality, solidity, and other such qualities. When a scientific controversy arises, scientists evaluate and discuss the reasons of the opponents; ``Scientific debate can continue indefinitely in search of a conclusion that can no longer be refuted" \citep[p. 6] {pievatolo_2024}. In contrast, in administrative evaluation of research, ``assessment authorities'' give evaluators decision-making power that allows them to decide when the debate should end and to impose their decisions ``even on those who disagree''\citep [p. 6] {pievatolo_2024}. 

In recent decades, administrative evaluation has become a ``taken-for-granted aspect of public administration and organizational procedures for management and development''\citep[p. 2]{dahler-larsen}. Hence, the administrative evaluation of research has also begun to be considered an inevitable aspect of research organization, despite the enormous variety of the ways in which it is used in different countries. This is the result of the widespread adoption of a consequentialist or utilitarian approach to administrative evaluation. In a nutshell, according to this vision, science has goals to achieve. These goals are defined by political authorities, supranational policies, or experts, perhaps after consultation with stakeholders. The scientific system is essentially designed as a system for producing science. The only relevant feature of the scientific system is the efficiency of production; hence, socially desirable scientific institutions are efficient institutions. The allocation of resources to these institutions should aim to maximize the expected outcomes. Administrative research evaluation is functional in enhancing the efficiency of research institutions; it allows political authorities or governing bodies to gather information to steer research institutions at distance. In particular, it allows one to properly allocate resources to the institutions that will be able to make the best use of them. In turn, this mechanism generates a continuous push toward improving research, enhancing the efficiency of the research system, and the growth of the scientific wealth of nations \citep{Whitley, dahler-larsen, Davis, Hammarfelt}. This approach is fully defined inside a model that argues that the quality and efficiency of science institutions are improved by introducing management techniques and practices that mimic the functioning of the market. According to  \citet{Bleiklie}, this model was the result of the recipes of the new public management, and it can be labelled as ``neoliberal'' (see also \citep{lave, Mirowski,Kulczycki}).

In fact, the idea that administrative evaluation of research can be taken-for-granted has found much resistance among scientists, who have questioned not only its effectiveness (for a recent overview see \citep{Schweiger_et_al_2024} and the bibliography cited therein), but also its legitimacy and neutrality with respect to the goals of science (see, among others, \citep{gillies, Collini_2012, Taylor_2022}).

\section{CoARA and the promotion of evaluative technocracy}

In March 2021 the European Commission (EC), Directorate for Research and Innovation, Unit A.4 Open Science, started a consultation with ``European and international stakeholders on how to facilitate and speed up reform so that the quality, performance and impact of research and researchers are assessed on the basis of more appropriate criteria and processes'' \citep[p. 3]{EC_scoping_report}. The EC considered that the current research assessment system "uses inappropriate and narrow methods to assess the quality, performance and impact of research and researchers. Notably, the quantity of publications in journals with high
Journal Impact Factor and citations are currently the dominant proxies for quality, performance and impact". Consultation results were synthesized in a scoping report published in November 2021. It  proposed the writing ``of a European agreement that would be signed by individual research funding organisations, research performing organisations and national/regional assessment authorities and agencies, as well as by their associations,
all willing to reform the current research assessment system'' \citep[p. 3]{EC_scoping_report}. The text of the Agreement on Reforming Research Assessment (ARRA) was written from February to July 2022. The writing of the ARRA was conducted by a drafting team of 12 people from European Commission and Science Europe, a non-profit international organization gathering forty national research funding organizations. A group of 19 ``contributors'' participated in the writing through comments and suggestions \citep{ARRA}. The draft of ARRA was also discussed in three assemblies involving organizations that had participated in the consultations for the scoping report. Finally, the member states of the European Union were also consulted (for a more detailed reconstruction of the whole process, see \cite{di_donato}). On 28 September 2022 the agreement was officially opened for signature and CoARA, the Coalition for the Advancement of Research Assessment, was launched.  CoARA is the ``collective organization'' in charge of putting the ARRA principles and commitments into practice. The modalities of governance, operations and funding of CoARA was drafted by ``representatives from the European University Association, Science Europe and the European Commission'' with the advice of an implementation group ``made of representative'' from ten organizations \citep{coara_organiz}. On December 1st 2022 CoARA was officially launched during a ``Constitutive assembly'' that adopted the governance document. Actually, CoARA's Secretariat is hosted by the European Science Foundation,  a non-governmental, internationally-oriented, non-profit association established in France in 1974. It coordinates also the CoARA Boost Horizon Europe project, a project financed with 5 million Euros by the European Commission \citep{coara_boost}.  

The European Commission presented ARRA as ``a co-creation process involving more than 350 organizations from 40 countries'' \citep{EC_2022}. In a recent European Union report the whole process is summarized as ``the [European] Commission facilitated a co-creation process to reach an agreement on reforming research assessment'' \citep[p. 13]{RRA}. Matthias Koenig, one of the current member of the steering board, even describes CoARA as ``a science-driven, bottom-up and inclusive coalition'' \citep{coara_board}.

However, despite the appearance of broad consultation and stakeholder involvement, the key orientations, procedures, and institutional architecture of CoARA were all designed and directed by the European Commission, with the support of selected ``representative organizations''. Notably, even ``representative organizations'' and the stakeholders involved in the consultation were possibly chosen by the very institutions responsible for drafting the agreement. The rationale on the basis of which drafters, contributors, organisations and stakeholders are selected is not disclosed, nor is the information on from whom they receive mandates, funding or reimbursement of expenses, to whom they are accountable for decisions made \citep{Pinto}. In sum, the text of ARRA, the governance structure, and operational modalities of CoARA were predefined by a restricted group of institutional actors ``facilitated'' by EC. Walter Rosenthal, the Vice president of the German Rector's Conference, stated syntetically that CoARA ``is basically the result of a top-down process'' \citep{Naujokaityt}.

As suggested in the scoping report, CoARA agreement is exclusively addressed to ``research funding organizations, research performing organizations, national/regional assessment authorities and agencies'' \citep[p.2 ]{CoARA2024}.  Once an organization has signed the agreement or joined CoARA as a member, signatory institutions and members of CoARA are granted operational flexibility in how they implement the commitments undertaken: organization have to present an action plan within a year of signing the agreement, and they ``have full freedom in the develoment of their Action Plan'' \citep{coara_action}. At April 2025, 191 action plans are available in Zenodo repository (https://coara.eu/agreement/action-plan/). 

The responsibility for translating CoARA’s principles into concrete practices rests entirely with the member institutions and organization. Indeed, CoARA does not directly engage with individual researchers: for example, it does not require them to sign any form of ethical or deontological commitment. Researchers's contribution to reform consists only of supporting the dissemination and promoting the adoption of COARA principles within their institutions. In this respect, CoARA differs from DORA that is aimed at a broader target audience, including researchers who, by signing its commitments, agree to act in a deontological manner, adopting specific behaviors in research and when acting as evaluators \citep[see recommendations 15-18]{dora}.

The domain of CoARA is solely ``research assessment'', the definition of which coincides with that of administrative evaluation of research. According to CoARA, ``research assessment'' encompasses ``the assessment of research performing organizations and research units, by assessment authorities, research funding and performing organizations; [...] The assessment of research projects by assessment authorities, research funding and performing organizations, and prize awarding organizations; [...] The assessment of individual researchers and research teams by research funding and performing organizations and prize awarding organizations'' \citep{CoARA2024}.

CoARA considers administrative evaluation of research as a taken-for-granted aspect of the contemporary organization of science, by avoiding the discussion about its legitimacy and neutrality with respect to the goal of science. More precisely, CoARA explicitly assumes that administrative evaluation of research is desirable for science and society: ``the assessment of research, researchers and research organizations recognizes the diverse outputs, practices and activities that maximize the quality and impact of research'' \citep[p. 1]{CoARA2024}. CoARA contributes to the administrative evaluation of research by simply adding that the adoption of evaluations techniques coherent with the principles of the agreement is essential for efficiency, i.e. to ``maximize the quality and impact of research''. According to CoARA, ``peer review is the most robust method known for assessing quality and has the advantage that it is in the hands of the research community'' \citep[p. 5]{CoARA2024}.

Indeed, CoARA's four core commitments concern the technologies to be adopted or avoided for administrative evaluation. The first consists in the recognition in administrative evaluation of the diverse outputs, practices, and activities that reflect the variety of contributions to science. The second requires that administrative evaluation be based ``on qualitative evaluation for which peer review is central, supported by responsible use of quantitative indicators''. The third commitment supports the abandonment of ``inappropriate uses of journal- and publication-based metrics, in particular inappropriate uses of Journal Impact Factor (JIF) and h-index''. The fourth commits the signatories of the agreement to ``avoid the use of rankings of research organizations in research assessment'' \citep[pp. 4-6]{CoARA2024}. 

The other six supporting commitments ``include three commitments to enable the move towards new research assessment criteria, tools and processes, and three commitments to facilitate mutual learning, communicate progress and ensure that new approaches are evidence informed'' \citep[p. 7]{CoARA2024}.

It is worthwhile to note that CoARA, particularly its second and third commitments, does not incorporate specific actions such as those proposed in DORA \citep{dora} or in the Leiden Manifesto for research metrics \citep{leiden}. Both documents are explicitly referenced only in Annex 4 of the Agreement, but the annexes “do not form an integral part of the Agreement” and are “provided to support its implementation” \citep[p. 12]{CoARA2024}. Moreover, as noted above, the Agreement grants signatories full discretion in selecting the tools and approaches to be adopted in their action plans. As a result, these plans may diverge significantly from the specific principles laid out in DORA and the Leiden Manifesto (see for instance \citep{pievatolo_anvur}). This discretion also extends to open science and open access, which are neither mandated nor operationalized in a uniform or binding way. The definition of “openness” provided in the Agreement is rather narrow: “openness corresponds to early knowledge and data sharing, as well as open collaboration including societal engagement where appropriate” \citep[p. 3]{CoARA2024}. Signatories are only committed to rewarding “research behaviour underpinning open science practices such as early knowledge and data sharing as well as open collaboration” \citep[p. 4]{CoARA2024}. In contrast, the broader and more robust principles of the UNESCO recommendation on open science \citep{UNESCO} are relegated to Annex 4, while open access is mentioned only in Annex 1. Given this framework, it is perhaps unsurprising that action plans might be developed with little or no attention to open science or open access \citep{Galimberti}. 

In general, the ten commitments adopt an `internalist' or technological' approach to administrative evaluation of research and its reform. In a nutshell, the reform of the administrative evaluation of research proposed by CoARA consists in a change of technology for evaluation. It is possible to express this point with the metaphor of the thermometer. At least since \cite{moed}, bibliometrics is represented as a thermometer, i.e. the (best) technology used for quantitative research evaluation. CoARA proposes to replace this technology, the bibliometric thermometer, with another one, the peer review, supported by the responsible use of quantitative indicators. 

CoARA implicitly argues that the definition of the most appropriate evaluation technique can be developed independently of its institutional setting, of its procedures and of the incentive system associated with the administrative evaluation of research. In essence, CoARA implicitly assumes that the choice of evaluation techniques can be made without considering the research policy framework in which the administrative evaluation of research is performed. In other words, CoARA refers to very different evaluation contexts at the same time, such as the massive evaluation of research in a PBRF system such as REF and VQR, or procedures for hiring a post-doctoral researcher. The only contextual element explicitly considered by CoARA consists in acknowledging that appropriate techniques must be adapted to the level of granularity of the evaluation (individuals, groups and institutions). 

CoARA lacks of a precise definition of ``peer review'' and a discussion of the many types of ``peer review'' adopted in contemporary science. It appears that CoARA promotes the adoption of qualitative evaluations by the peer community, regardless of the specific ways in which they are performed. The replacement of metrics with peer review assumes, in fact, that when researchers act as evaluators, they are, as scientists, naturally able to recognize research quality and impact. CoARA admits that evaluators can be consciously or unconsciously biased on gender, ethnic origin, and other, and encourages training evaluators to recognize and mitigate biases \citep[p. 22]{CoARA2024}. It is implicit in CoARA that ``unbiased'' evaluators are committed to the good of science, independently of the institutional and administrative framework in which the assessment is realized. This appears hard to defend, as it asks scientists not to be self-interested. A more realistic way of representing the behavior of scientists engaged in administrative evaluation consists in admitting that they may act as ``rational referees, who might not have any incentive to see high-quality work other than their own'' \citep{Thurner} or of their group or institution. This is all the more true since the issues involved are mostly about money (e.g., hiring, promotions, funding). Agent-based simulations on peer-review in journals show that the presence of a minority of rational or selfish referees is sufficient to drastically lower the quality of the scientific standard \citep{Thurner, Cabota}. CoARA appears to hide or ignore such problems. 

This replacement of technologies also assumes that there are no differences between the evaluation of research carried out within an administrative process, such as the hiring of a researcher or the national research evaluation, and the common evaluative practices adopted by scholarly communities to recognize scientific claims as provisionally true \citep{Kitcher,pievatolo_2024}. More precisely, CoARA operates in the slippery landscape as defined by the argument that peer review ``is in the hands of the research community'' \citep[p. 5]{CoARA2024}: the judgments of scientists involved in the administrative evaluation of research are of the same nature or type as those produced in the peer review processes of academic journals \citep{Bornmann_2013}, and as those discussed by scientists in public debates and controversies about science. In fact, all these evaluative practices use different information, possibly adopt different standards, and happen in different contexts. COARA hides or ignores such differences.

In summary, CoARA designs the efficient organization of research as an evaluative technocracy. In fact, CoARA assumes that administrative evaluation of research is desirable for science and society. The path to reforming the administrative evaluation of research requires that organizations, authorities, and agencies in charge of it adopt the ten CoARA commitments.  These commitments are focused on the technology to be adopted in the administrative evaluation of research: peer review supported by the responsible use of quantitative indicators. The adoption of peer review requires that the evaluation be carried out by researchers whose judgments have administrative value and are not subject to public discussion, but possibly only to legal challenge.
 
\section{Philip Kitcher's well-ordered science}

In order to frame the role of administrative evaluation of research and of CoARA in the current organization of science, I suggest to take seriously Philip Kitcher's ideas of science in a democratic society\citep{Kitcher}. Kitcher adopted John Rawls' theory of justice \citep{rawls} and developed its application to the system of science. Just as Rawls outlines the characteristics of a well-ordered democratic society, Kitcher defines an ideal system of science, which he labels ``well-ordered science'', compatible with the institutions of a Rawlsian democratic society. 

Kitcher defines science as a public knowledge system in which researchers contribute new discoveries. The public knowledge system is organized in four sequential processes: investigation, submission, certification and transmission of knowledge, ``elaborated in different ways in different societies'' \citep{Kitcher}. Each process poses a series of questions that Kitcher aimed to respond by avoiding both ``pathologies of direct democratic control of science and an expert-driven technocracy that separates science from society'' \citep{sep-scientific-pluralism}.

For the investigation process, the main question is to define what ``relevant'' scientific inquiries should be pursued in a democratic society. Who has the authority to decide what are the relevant problems that science should deal with? In well-ordered science, the specification of ``the problems to be pursued would be endorsed by an ideal conversation, embodying all human points of view, under conditions of mutual engagement''.  Participants in this ideal conversation ``are to have a wide understanding of the various lines of research, what they might accomplish, how various findings would affect others, how those others adjust their starting preferences, and the conversationalists are dedicated to promoting the wishes other participants eventually form'' \citep{Kitcher}. It is easy to recognize the similarity with individuals in ``original position'' participating to the  definition of the principles of justices in Rawls' \textit{Theory of Justice} \citep{rawls}.

The introduction of the ideal discussion allows Kitcher to individuate two emerging dangers in defining the relevant issues to be investigated. According to Kitcher, in the well-ordered science of a democratic society, no one should have the authority to decide what the relevant issues are (for a Kantian perspective see \cite{pievatolo_2024}). The relevant issues cannot be identified by a simple majority voting procedure, because this could lead to the emergence of the ``tyranny of ignorance'', where myopic voters choose ``in ignorance of the possibilities, and of the consequences for others, completely absorbed in their own self-directed wishes". The second danger is the ``tyranny of unwarranted expertise'' that occurs if researchers are given the right not only to do research, but also to choose which topics should be investigated. Well-ordered science takes care to avoid these two tyrannies in the definitions of relevant issues to be investigated. 

According to Kitcher, the processes of submission and certification refer to the public discussion to which the results of science are subjected. Submission poses the questions of ``which people are entitled to submit reports to the public depository. [...] How are they trained? What standards do we expect them to meet in their investigations?'' \citep{Kitcher}. Certification is the process through which new ideas are provisionally accepted or provisionally rejected as, respectively, valid or invalid part of the public knowledge. Here, questions arise about what is required to accept or reject proposals as part of the body of public knowledge. The usual submission of results to ``agencies of the public knowledge systems, typically journals and other vehicles of publication'', the review process and the possible acceptance for publication are but the first phase of certification. It may happen that after initial certification of a report, subsequent works lead to its correction or even to its retraction. `` For a submission to be certified in the fullest sense, a community of inquirers must count it as true enough and important enough'' \citep{Kitcher}. 

The certification procedure must also be subject to scrutiny by outsiders, a condition that Kitcher calls \textit{ideal transparency}. ``A system of public knowledge is ideally transparent just in case all people, outsiders as well as researchers, can recognize the methods, procedures and judgments used in certification (whether they lead to acceptance or rejection of new submissions) and can accept those methods, procedures, and judgments'' \citep{Kitcher}. In sum, a well-ordered certification requires an ideal deliberation consisting of a public discussion involving researchers and citizens. These conditions guarantee that the public knowledge system is a collective undertaking endorsable by all, storing sufficiently meaningful and truthful information.

Certification itself is subject to the two dangers of the tyranny of expertise and the tyranny of ignorance. The first consists in the prevalence within scientific subcommunities of pervasive ideological agendas or biases that drive members to favor particular hypotheses, overvalue certain types of evidence, or fabricate evidence. Pervasive misjudgment, according to Kitcher, is particularly relevant to the current erosion of scientific authority, in turn connected to the growing body of results based on manipulation and fraud insinuating into the accepted corpus of scientific findings. Contemporary science is far from a generalized adoption of Mertonian norms \citep{merton}. Hence, both submission and acceptance procedures and replication alone do not suffice to detect and deter fraud. 

The overreaction to the tyranny of expertise led to the tyranny of ignorance. Kitcher refers here to Paul Feyerabend's call for ``vulgar democracy'' directly in the context of certification: ``duly elected committees of laymen'' are called to deliberate by majority voting if scientific results are really well established and should be considered part of the public knowledge worth passing in textbooks for future generations \citep{feyerabend}. The ``mob rule'' defended by Feyerabend is but a ``vulgar surrogate'' of the ideal deliberators of well-ordered certification.

Finally, the process of transmission involves questions about what parts of the public knowledge are available to whom and whether the knowledge needed by different people is transmitted to them. Serious shortcomings in access to information interfere, ``possibly disastrously'', with the functioning of democratic societies. These questions are intertwined with the functioning of education in society.

\section{Administrative evaluation of research and well-ordered science}

Kitcher's highly idealized model of public deliberation in well-ordered science has had limited appeal to science governance scholars \citep{sep-scientific-pluralism}. However, it brings to reflect on the role and positioning of the administrative evaluation of research of individuals, projects, and institutions. The question to answer what the role is of administrative evaluation of research in a Kitcher-like logic.  

To answer the question, one can start from the observation that in administrative procedures of research evaluation, evaluators are academics with administrative authority. Consider, for example, the evaluation of a research article in a massive research assessment. This evaluation is neither the result of a well-ordered certification nor the result of a public discussion among peers. It is instead the indisputable judgment of one or more scientists invested of the administrative authority to act as evaluators. As highlighted by \cite{pievatolo_2024}, the fact that evaluators are scientists does not signify that they are `peer' of the scientists being evaluated, since only the former have administrative authority. The evaluated scientists can challenge the judgment they receive not through reasoning in public debate but only through administrative or judicial action. In Italy, for example, to challenge the judgment received in the national research assessment (VQR) or in the procedures for the qualification (ASN), a researcher must resort to a court action before an administrative court. In short, evaluators act as 'guardians of science' for the policy makers who have granted them the authority to evaluate.

Given this characterization, it can be argued that administrative evaluation of research does not produce a Kitcher-like orderly certification, for two reasons. First, the research products to be evaluated are already part of the body of provisionally accepted public knowledge and are therefore already provisionally certified: discussion about them has taken place and is ongoing because they are publications that, as such, are subject to the judgment of peers and the reading public. The second reason is procedural: the outcome of the evaluation of individual research products is not only not debatable, but not even normally made public. This formally excludes it from being a certification.

Therefore, administrative evaluation of research can be properly characterized as a value-laden post-certification administrative activity. The adjective ``administrative'' indicates that it takes places under the umbrella of an organization, institution or agency vested with the power and responsibility to manage and regulate the evaluation. The expression ``value-laden'' indicates that evaluation happens according to criteria and standard defined directly by a 'government' or by institutions, agencies or individual researchers vested with the power of evaluation. The fact that these criteria mimic those used in peer review of scholarly journals and that they are applied by academics does not give the evaluation a Kitcher-like certification character. ``Post-certification'' indicates that research products to be evaluated usually are already published and as such provisionally certified. Moreover, as already highlighted, the authority of the evaluators ultimately derives from a governmental act and not from participation in the public conversation underlying the certification. 

Administrative evaluation of research is subject to both the tyranny of ignorance and the tyranny of unwarranted expertise. To understand the point, a thought experiment can be proposed. Let us try to imagine a government authority deciding to give the power to evaluate the research articles produced by a university to a panel of individuals chosen at random from among the citizens. Criteria to be adopted are ``originality, significance and rigour'', as in the British REF. The panelists decide the final score of each article by majority vote. This would be a manifestation of the tyranny of ignorance, since lay people chosen at random from population are possibly unable to carefully judge ``originality, significance and rigour'' of research articles. 

Let us now imagine, alternatively, that the government authority chooses evaluators from among academics, that judgments are based on ``originality, significance and rigour'', and that the final score of each article is computed by more or less complex procedures (average of evaluations, majority vote or other).  In this case, paraphrasing Kitcher, it is not difficult to think of the emergence of the tyranny of unwarranted expertise: A particular ideological agenda, a pervasive bias, pushes members of a scientific subcommunity vested with the authority to evaluate to overestimate the quality and impact of certain products and to underestimate others even while fabricating the necessary evidence. For instance, evidence exist showing that the tyranny of unwarranted expertise was at work in British \citep{lee1998,lee2013} and Italian \citep{baccini_agreement, baccini_artifact, baccini_2025} research assessments.

In sum, administrative evaluation of research is at odds with Kitcher's well-ordered science for two main reasons. First, the definitions of quality and impact used for evaluation are not the result of a public discussion, but they are authoritatively stated by governments or institutions in charge of the administrative evaluation of research. Second, governments or institutions give a group of scientists the power to judge research based on these definitions. 

The administrative evaluation of research has a major effect on the functioning of research systems. Authoritarian definitions of quality and impact enter, more or less directly, in the design of research policies and of incentives for scientists, through PBRF systems, through recruitment, promotion, and career policies, or in the form of monetary rewards. In this way, they help to define the preferences of scientists as to which problems to investigate, which methods to use, and through which channels to disseminate research results. If scientists are toughts as rational individuals, they will produce research outputs and implement ``practices and activities that maximize the quality and impact of research'', just as CoARA advocates. And the quality and impact of research being maximized are not defined by the scholarly community, but by administrative authorities governing the procedures of research evaluation.

It is not unreasonable to think that this also has feedback on the criteria that the scientific community adopts for the certification of knowledge: quality and impact as defined by the authority in charge of the administrative evaluation of research can, in this way, replace the certification criteria previously constructed and adopted by the scientific communities. In this sense, administrative evaluation of research contributes to distancing more and more actual science from the ideal of well-ordered science.

\section{Discussion and conclusion}

The application of Kitcher’s concept of well-ordered science in the context of a democratic society allows for a clearer understanding of administrative research evaluation as an activity fundamentally external to the domain of science itself. Rather than serving as a scientific certification process, administrative evaluation functions as a mechanism for ascribing value to research outputs and contributions based on criteria established by administrative or policy authorities. These criteria are then implemented by scientists who, in this context, assume the role of administrative agents, rather than autonomous members of a scientific community. As such, the very act of implementing administrative evaluation shifts the practice of science away from Kitcher’s ideal of a well-ordered science, in which research is shaped by an ideal collective deliberation informed by both epistemic values and democratic input. In contrast, administrative evaluation imposes externally-defined priorities and bureaucratic standards, subject both to tyranny of ignorance and to tyranny of expertise.

The research assessment reform movement, and CoARA in particular, originates from the premise that administrative research evaluation is both necessary and inevitable. In the case of CoARA, this foundational assumption is reflected not only in its stated objectives but also in its institutional genesis. As discussed above, despite the appearance of broad consultation and stakeholder engagement, the initiative was largely orchestrated by the European Commission, with key orientations, procedures, and design features defined in advance by a limited group of actors. This continuity is not incidental. Since the Lisbon Strategy of 2000, when the European Union declared its ambition to become “the most competitive and dynamic knowledge-based economy in the world”, a succession of policy frameworks, including Europe 2020 and the renewed European Research Area (ERA), have institutionalized administrative research evaluation as a central governance mechanism. Now that these strategies appear to have difficulties to deliver on their transformative promises, CoARA can be understood as an effort to re-legitimize and adapt the evaluative apparatus to new discourses around openness, transparency, and inclusivity, without relinquishing its underlying logic.

The core problem CoARA seeks to address lies in the overreliance on quantitative metrics, such as citation counts and journal impact factors, which has led to widespread goal displacement: researchers optimize for what is measured rather than for what truly matters to scientific progress or societal benefit. The reform agenda is grounded in the belief that by adopting more appropriate, inclusive, and context-sensitive evaluative technologies is possible to realign research incentives and foster outcomes that are both scientifically meaningful and socially desirable. According to CoARA, the most promising technology consists of peer-review informed by the responsible use of metrics, that can be readily adapted to the diverse contexts and purposes of different evaluation procedures.

In turn, the notion that administrative research evaluation is necessary is also supported by scholars who advocate for the use of carefully constructed bibliometric indicators \citep[for all]{abramo2024}. In this perspective, the defense of informed peer review and that of bibliometrics are, in effect, two sides of the same coin. Both aim to enhance the administrative research evaluation, albeit through different means. This debate about alternative evaluative technologies is deeply embedded in a neo-liberal conception of science governance. Within this vision, research is organized and managed as a productive system whose outputs must be continuously monitored, evaluated, and optimized for ``maximise the quality and impact of research'' [Coalition for Advancing Research Assessment (CoARA), 2022, p. 1].

The two alternative proposals fail to recognize the authoritarian nature of administrative research evaluation as a structural problem within contemporary science. At its core, this model concentrates evaluative authority in the hands of a limited group of scholars, who are thereby empowered to “put an end to the free exchange of ideas and override the legitimate opinions” of other scientists \citep{Kitcher}.
CoARA’s commitments explicitly endorse the delegation of evaluative authority to selected academics deemed capable of conducting peer review, supported by the “responsible” use of metrics. Similarly, proponents of bibliometrics advocate for entrusting evaluation to expert bibliometricians \citep{abramo2024}. In this latter case, the technocratic character of research assessment is not only acknowledged but openly defended: evaluation must be governed by experts in quantitative indicators.
By contrast, CoARA obscures the authoritarian dimension of evaluation by framing it within the language of peer review, cloaked in ideals of collegiality and responsibility. Yet, in both cases, evaluative power remains exclusionary exposing research assessment to the risks of a tyranny of expertise, the removal of legitimacy from dissenting scientific perspectives and reinforcement of hierarchical control over what counts as valuable research.

From this perspective, the ``forced battle between peer review and scientometric research assessment'' \citep{abramo2024} does not appear to be an intellectual battle. Rather, it appears as a battle for the authority to evaluate, to occupy positions in government offices or agencies that conduct administrative evaluation of research, and to sell well-paid evaluative services. In fact, the growing space between policy makers and scientific system is progressively filled by academics transformed in professionals of evaluation, by consultancy agencies and data providers \citep{Jappe}. The winners n this battle over evaluation technologies will gain privileged access to the market for expertise they claim to possess, mirroring the trajectory of the illiberal reformers described by \cite{Leonard}. They turned technocratic authority into a platform for political and economic power. Leonard shows how, in the early 20th century, a new class of experts, economists, administrators, and policy advisors, used the language of science and rationality to displace democratic deliberation with technocratic control, often serving elite interests under the guise of neutrality. 

The obvious objection to the reasoning so far is that Kitcher's ideal conversation is only an impractical idealization, since the organization of contemporary science institutions necessarily requires administrative evaluation. The substitution of technology proposed by CoARA and the recommendations contained in the Leiden Manifesto could therefore be considered as a realistic and feasible proposals to reform administrative evaluation of research by shrinking or at least by correcting the use of metrics. 

The counter-objection begins by noting that the emergence of pervasive administrative research evaluation is a relatively recent phenomenon \citep{sandstrom}. Until the 1980s, national science systems operated without New Public Management mechanisms and without systematic administrative evaluation, relying instead on diverse organizational models shaped by specific historical and institutional contexts \citep{gillies, gingras, sandstrom, Whitley}. The exponential growth of contemporary science does not, in itself, justify the necessity of these administrative tools. As many critics have argued \citep{gillies, Collini_2012}, the problems that gave rise to the research assessment reform movement point not so much to the need for better evaluative technologies, but to the need for a more fundamental rethinking of the purpose, utility, and necessity of administrative research evaluation itself.

A possible option for research policy may be to abandon the precepts of new public management and to return to limit the administrative evaluation of research to a bare minimum. This minimum could consist in limiting the administrative evaluation of research to only those procedures strictly necessary for the basic functioning of science: the recruitment of researchers and the funding of project-based research. This means eliminating performance-based funding mechanisms for research institutions, massive individual and aggregate evaluation procedures, and direct cash incentives for publications and citations. The main advantage of this elimination is to greatly reduce not only the risks of tyranny of expertise and ignorance associated with, but also the distortions in scientists' behavior induced by administrative evaluation of research.

Even in a scenario where administrative research evaluation is strictly limited, the use of peer review, as proposed by CoARA, does not fully eliminate the risk of a tyranny of expertise. When such evaluation is employed for purposes like researcher recruitment or project funding, the central challenge becomes the design of \textit{fair} procedures \citep{beersma2003, baccini_2025}, aimed at minimizing the concentration of evaluative power and reducing the risk of epistemic exclusion. Procedural fairness emphasizes the integrity of the evaluation process itself, independently of its outcomes. It is a non-consequentialist notion: what makes a procedure fair is not the distribution of rewards it produces or its effects on research quality, but its internal characteristics, such as consistency, impartiality, and the inclusion of diverse perspectives and interests of those affected by the evaluation \citep{Leventhal1980}.

Even when administrative research evaluation is narrowly applied, as suggested here, and implemented through peer review, as proposed by CoARA, it still concentrates decision-making power in the hands of a panel, that is, a specific group of scholars. To minimize the risks of a tyranny of expertise, two conditions are crucial: ensuring fairness in the composition of evaluation panels and carefully limiting their authority \citep{beersma2003, baccini_2025}. For instance, selecting panel members should not rely solely on visible dimensions like gender or institutional affiliation, but must also reflect the plurality of epistemic approaches within a field \citep{baccini_2025}. In the allocation of research funding, fairness can also be enhanced by incorporating elements of randomization to mitigate bias and reduce the discretionary power of reviewers \citep{Avin_2018}. \citet{gillies} and others, for instance, have suggested, a two-stage process, where experts identify a pool of meritorious projects, from which final selections are made at random.

Ultimately, limiting the scope of administrative research evaluation and designing fair evaluative procedures can help mitigate both the tyranny of expertise and the tyranny of ignorance in contemporary science systems. This is a goal that CoARA, and, more broadly, any reform agenda focused exclusively on improving evaluative technologies, cannot achieve on its own, unless it is accompanied by a more fundamental rethinking of the role, purpose, and governance of research evaluation itself.

\newpage

\bibliography{references}  






\end{document}